# Quantum critical point and spin fluctuations in the lower-mantle ferropericlase

I. S. Lyubutin[1], V.V. Struzhkin[2], A. A. Mironovich[3], A. G. Gavriliuk[1-3], P. G. Naumov[1], J. F. Lin[4], S. G. Ovchinnikov[5,6], S. Sinogeikin[7], P. Chow[7], and Y. Xiao[7]

[1]*Institute of Crystallography, Russian Academy of Sciences, Moscow 119333, Russia*

[2]*Geophysical Laboratory, Carnegie Institute of Washington, Washington DC, USA*

[3]*Institute for Nuclear Research, Russian Academy of Sciences, Troitsk, Moscow 142190, Russia*

[4]*Department of Geological Sciences, Jackson School of Geosciences, The University of Texas at Austin, Austin, Texas 78712-0254*

[5]*L.V. Kirensky Institute of Physics, SBRAS, Krasnoyarsk, 660036, Russia*

[6]*Siberian Federal University, Krasnoyarsk, 660062, Russia*

[7]*Advanced Photon Source, Argonne National Laboratory, Argonne, USA*

# ABSTRACT

Ferropericlase, (Mg,Fe)O is one of the most abundant minerals of the Earth's lower mantle. The high-spin (HS) to low-spin (LS) transition in the $Fe^{2+}$ ions can dramatically alter the physical and chemical properties of (Mg,Fe)O in the deep mantle, thereby changing our understanding of the Earth's deep interior. To establish a fundamental understanding of the ground electronic state of iron, the electronic and magnetic states of $Fe^{2+}$ in $(Mg_{0.75},Fe_{0.25})O$ have been investigated by transmission (TMS) and synchrotron (NFS) Mössbauer spectroscopy at high pressures and low temperatures (down to 5 K). The results show that the ground electronic state of $Fe^{2+}$ at the critical pressure $P_c$ of the spin transition and close to T=0 is determined by a quantum critical point $P_q$ (T = 0, $P_c$) where the energy difference between the HS and LS states (an energy gap for the spin fluctuation) is zero. The deviation from T=0 leads to the thermal excitation for the HS or LS state, suggesting a strong influence on the magnetic and hence the physical properties of the material. Combining these with theoretical calculations, the results indicate that the existence of the quantum critical point at zero temperature affects not only the low-temperature physical properties, but also the strong temperature/pressure-dependent properties at conditions relevant to the middle layer of the lower mantle.

## 1. INTRODUCTION

Ferropericlase (Mg,Fe)O in the face-centered cubic rock-salt structure is believed to be the second-most abundant mineral phase in the Earth's lower mantle (approximately 30% abundance by volume with 20 atomic % of Fe) after silicate perovskite (Mg,Fe)SiO$_3$ with approximately 10 atomic % Fe. [1-6] Great interest in this system has been attributed to the pressure-induced electronic transition in (Mg,Fe)O, in which the Fe$^{2+}$ ions transform from the high-spin (HS) state ($S = 2$) to the low-spin (LS) state ($S = 0$). [7-14] A series of physical and chemical properties of (Mg,Fe)O can be dramatically altered by the spin transition in the deep Earth's mantle, including thermal [15, 16] and electrical [17] conductivities, density, [18, 19] compressibility, [18] and sound velocities, [20] among others. Most previous investigations on the spin transition of iron in ferropericlase were performed at high pressures and room to high temperatures, where the Fe ions in the starting samples were in the paramagnetic high-spin state and/or diamagnetic LS state. Studies on the electronic and magnetic states at low temperatures would provide critical information on the ground state of the Fe ions in ferropericlase at high pressures. [21] Here we have investigated electronic and magnetic properties of Fe$^{2+}$ in two representative compositions of ferropericlase (Mg$_{1-x}$Fe$_x$)O (x=0.25 and 0.2) at high pressures and low temperatures using transmission and synchrotron Mössbauer spectroscopy in diamond anvil cells (DACs) up to 90 GPa. Based on the analyses of the Mössbauer spectra, the derived hyperfine parameters of iron ions in the sample are used to construct the magnetic phase diagram and to address the quantum critical point phenomenon in (Mg,Fe)O at high pressures and low temperatures where the spin gap energy between the HS and LS states is zero. Based on the theory of the quantum spin fluctuations, we predict an appearance of new magnetic properties in (Mg,Fe)O at the high P-T conditions relevant to the Earth's lower-mantle.

## 2. EXPERIMENTAL TECHNIQUE

Polycrystalline (Mg$_{0.75}$,Fe$_{0.25}$)O and (Mg$_{0.8}$,Fe$_{0.2}$)O samples with 95% or 20% $^{57}$Fe enrichment, respectively, were synthesized by the ceramic method.[7] X-ray diffraction patterns showed that the samples had the rock-salt structure, whereas neither magnetite ($Fe_3O_4$, in the spinel structure) nor hematite ($\alpha$-$Fe_2O_3$, in corundum structure) were detected in the X-ray diffraction patterns. The $^{57}Fe$ Mössbauer spectral analyses revealed the synthesized $(Mg_{0.8},Fe_{0.2})O$ and $(Mg_{0.75},Fe_{0.25})O$ samples contained approximately 2% and 7% of the ferric $Fe^{3+}$ iron ($Fe^{3+}/(Fe^{2+}+Fe^{3+})$), respectively.[14][13] The small amount of ferric $Fe^{3+}$ ions may be localized either in interstitial sites or in substitution for the octahedral Mg sites, forming dimers with cation vacancies (□) in order to balance the molecular electro-neutrality [22] ($Fe^{3+}$ — □ — $Fe^{3+}$) = $3Fe^{2+}$.

At ambient pressure, the $^{57}Fe$-Mössbauer spectra were recorded in the temperature range of 4.2 K to 300 K in the transmission geometry with a standard spectrometer operating in the constant accelerations regime (the TMS technique). The gamma-ray source $^{57}Co(Rh)$ was used at room temperature, and the isomer shifts were measured relative to $\alpha$-Fe metal at room temperature. Average values of the magnetic hyperfine field $<H_{hf}>$ at iron nuclei were estimated from the distribution functions of the hyperfine parameters constructing for widened spectral lines.

High-pressure synchrotron Mössbauer spectroscopic technique [nuclear resonant forward scattering (NFS)] was applied to study magnetic and electronic spin states of iron ions in $(Mg_{0.75},Fe_{0.25})O$ at pressures of up to 90 GPa and temperatures ranging from 8 K to 300 K. For the NFS measurements, powder $(Mg_{0.75},Fe_{0.25})O$ samples were flattened down to approximately 3 μm thick disks between two diamond anvils. Rhenium gaskets were pre-indented to a thickness of 25 μm and a hole of 80 μm was drilled in them. Small disks of the samples with a size ~ 50 x50 μm$^2$ were loaded into the sample chambers of the DACs with flat diamonds of 300 μm in culet size. Helium gas was loaded into the sample chambers as the pressure medium. The synchrotron beam-size was focused to a small spot of about 7x7 μm$^2$ which allowed us to collect all NFS spectra over an extremely small sample area that provided nearly hydrostatic conditions with negligible pressure gradients. Several ruby chips, each of 1-2 μm in diameter, were placed next to the samples in the chambers for in-situ pressure determinations and evaluations of the pressure gradients, which showed that the pressure gradients of the measured sample areas did not exceed 2 GPa at the maximum experimental pressure of 90 GPa.

A specially-designed DAC [23] and small cryostat have been used for the low-temperatures measurements, in which pressure can be held stable during cooling process to 4.2 K. Temperatures of the sample chambers were controlled within ±2 K using a feedback power supply unit. The NFS measurements in DAC were carried out at the beamline 16-IDD of the Advanced Photon Source (APS), Argonne National Laboratory (ANL). [7] The measured NFS spectra were analyzed with the MOTIF program [24].

## 3. EXPERIMENTAL RESULTS AND DISCUSSIONS

### 3.1. Low-temperature Mössbauer data at ambient pressure

Low-temperature, ambient pressure TMS spectra are shown in Fig. 1(a) for $(Mg_{0.75},Fe_{0.25})O$ and Fig. 1(b) for $(Mg_{0.8},Fe_{0.2})O$. At 5 K, the spectral lines are split by the magnetic hyperfine interaction, indicating the magnetic ordering of $Fe^{2+}$ ions. The magnetic splitting decreases gradually with increasing temperature, whereas the spectral features are dominated by a doublet typical of the paramagnetic state above 40 K. [14] In the magnetic region, the spectral lines are very broad and asymmetric. The line broadening can be explained by the presence of many non-equivalently-distributed $Fe^{2+}$ sites resultant of nearest neighbor interactions. As discussed in details recently, [14] a large number of nonequivalent iron sites differing in numbers of Fe and Mg ions in the nearest and next-nearest neighboring iron ion positions have been observed in $(Mg_{1-x},Fe_x)O$. On the other hand, the line asymmetry can be explained by the higher electric quadrupole interaction, which becomes comparable with the magnetic hyperfine interaction at temperatures near the Néel temperature ($T_N$). [25]

By plotting the distribution functions for the hyperfine parameters, an estimation of the averaged value of the magnetic hyperfine field at the iron nuclei $<H_{hf}>$ can be determined. The temperature-dependent $<H_{hf}>$ values are shown in Fig. 1 (c) and (d) for $(Mg_{0.75},Fe_{0.25})O$ and $(Mg_{0.8},Fe_{0.2})O$, respectively. Based on this behavior, we have also evaluated the Néel temperatures $T_N$ as ~37 K and ~27 K for $(Mg_{0.75},Fe_{0.25})O$ and $(Mg_{0.8},Fe_{0.2})O$, respectively. The $T_N$ value for $(Mg_{0.8},Fe_{0.2})O$ was found to be close to that given by Speziale *et al.* at 25 K. [21] It is known that pure wüstite (FeO), an endmember of the $(Mg_{1-x},Fe_x O)$ series, is an antiferromagnet with the $T_N$ of about 198 K, and its crystal structure changes from cubic to rhombohedral upon the magnetic transition at high pressures. [6, 26] The magnetic ordering is mainly governed by strong super-exchange antiferromagnetic interactions in Fe-O-Fe with a bond angle of about 180º. Similarly, the magnetic properties of the $(Mg_{1-x},Fe_x)O$ solid solution depend on the iron concentration ($x$) and can also be determined from the Fe-O-Fe super-exchange interactions. Due to percolation processes, long-range magnetic ordering may appear at certain iron concentrations above the critical value $x_c$. For a three dimensional *fcc* lattice, the critical value is close to $x_c \approx 0.16$. [27] Thus both compounds studied here have iron concentrations above the percolation limit and can exhibit a long-range magnetic order below $T_N$.

We have also investigated the behavior of the quadrupole splitting ($\varepsilon$), the isomer shit ($\delta$, the central gravity shift), and the area of the resonance lines ($I$) in the paramagnetic state at temperatures between $T_N$ and 300 K. As shown in Fig. 2, the average value of the quadrupole splitting in $(Mg_{0.75},Fe_{0.25})O$ and $(Mg_{0.8},Fe_{0.2})O$ decreases continuously from about 1.90 to 0.75 mm/s in the temperature range, whereas the isomer shift decreases in accordance with the relativistic second-order Doppler effect (the temperature shift). [25] Considering that the area $I$ of the resonance lines is proportional to the probability of the Mössbauer effect ($f'$), we fit the experimentally-derived $I$ values to the Debye approximation, [28] and evaluated values of the "Mössbauer" Debye temperatures ($\theta_D$) for the local $Fe^{2+}$ sites in $(Mg_{0.75},Fe_{0.25})O$ and $(Mg_{0.8},Fe_{0.2})O$ are 290 (±10) K and 450 (±10) K, respectively.

### 3.2. Synchrotron Mössbauer data at high pressures and low temperatures

The NFS spectra of $(Mg_{0.75},Fe_{0.25})O$ were recorded at high pressures up to 90 GPa and at temperatures between 8 and 300 K (Fig. 3a and 3b). Contrary to the TMS technique where the resonance signal is recorded as function of energy of the Mössbauer gamma-quanta, the NFS signal is a function of time. The time spectra represent a damped decay of the nuclear excitation which is modulated in time by quantum and dynamic beats (see details in [24, 29]). At lower temperatures, 12 and 15 K, the high-frequency quantum beats of the magnetic signature are present in the spectra (Fig. 3a), indicating the occurrence of the magnetic ordering of the iron ions. Above 50 K, only low-frequency quantum beats can be seen in the NFS spectra revealing the paramagnetic state of (Fig. 3a). Analyses of the shapes of the quantum beats show that the samples transform from a low-frequency paramagnetic state to a high-frequency magnetic state at pressures below 56 GPa during temperature cooling cycles (Fig. 3a). The low-frequency quantum beats appear due to the electric quadrupole interaction of the $^{57}$Fe nuclei with the electric field gradient (EFG) at the local iron sites. Disappearance of the magnetic quantum beats at $P < 56$ GPa can be considered as the transition from the magnetically-ordered state to the paramagnetic state (both are in the HS states). Based on the analyses of the NFS spectra, we have derived the Néel temperatures of the samples at pressures below 56 GPa. It should be noted that the magnetic transitions of ferropericlase $(Mg_{0.75},Fe_{0.25})O$ can be rather complicated at temperatures below the $T_N$ because of the existence of the many nonequivalent iron sites [14] and the percolation effects in the magnetically-diluted system. At pressures above 56 GPa, the low-frequency quantum beats of the electric quadrupole signature have disappeared at T < 50 K, and the NFS spectra of the nuclei decay appear as straight lines (Fig. 3b). This spectral shape corresponds to a singlet in the TMS spectra with zero quadrupole splitting, [14] showing the occurrence of the LS state of $Fe^{2+}$.

Above 90 K, the low-frequency quantum beats have appeared in the spectra showing the presence of the paramagnetic HS state with increased abundance with rising temperatures (Fig. 3b). Near the critical pressure of 56 GPa, we have observed the coexistence and fluctuations of the HS and LS $Fe^{2+}$ irons with relative abundance changing dramatically with temperature over the spin gap. That is, we have observed the thermal activation effect on the iron spin states at extreme conditions. This effect can be understood in terms of the thermal fluctuations through a spin gap. Near the critical pressure, the spin-gap energy $\varepsilon_S$ is

$$\varepsilon_S = E_{HS}\left(d^6, {}^5T_2\right) - E_{LS}\left(d^6, {}^1A_1\right), \tag{1}$$

where the $\varepsilon_S$ value is proportional to $(P - P_c)$. As will be shown below, at $P > P_c$ the thermo-activated magnetic moment may appear (from the HS $Fe^{2+}$ ions) with a maximum magnetization at $T_S \sim \varepsilon_S$ as a result of this process.

### 3.3. Magnetic phase diagram of $(Mg_{0.75},Fe_{0.25})O$ and a quantum critical point

Based on the analyses of our Mössbauer data at high pressures and low temperatures, a magnetic phase diagram of $(Mg_{0.75},Fe_{0.25})O$ is reported in Fig. 4. The diagram shows regions of the HS paramagnetic and antiferromagnetic phases at 0-50 GPa, and the LS diamagnetic phase above 56 GPa. The pressure-dependent Néel temperatures (the dash blue line in Fig. 4) in the region $0 < P < 50$ GPa separate the magnetically-ordered and paramagnetic HS states of $(Mg_{0.75},Fe_{0.25})O$. The $T_N$ value increases with increasing pressure and reaches the maximum value of about 55 K at 30 GPa, but drops dramatically as the critical point of the HS-LS crossover was approached. At the absolute zero temperature and the critical pressure of 56 (±3) GPa, a quantum critical point appears in the magnetic phase diagram.

At T=0 and $P = P_c$, the physical meaning of the quantum critical point can be clarified below. At T = 0 and $P < P_c$, the ground state of the $Fe^{2+}$ ion in the HS state has the wave function $\Psi(HS)$. At each pressure in the $P < P_c$ range, the LS state is separated from the HS state by the energy $|\varepsilon_S|$, and can be populated by thermal excitations only. Above $P_c$ at T = 0, the ground state of $Fe^{2+}$ is low spin and has the wave function $\Psi(LS)$. The HS state is separated from the LS state by the energy $|\varepsilon_S|$ and can also be populated only by thermal excitation. In the quantum critical point (T = 0, $P = P_c$) the energies of the HS and LS states are equal. The wave function for the $d^6$ configuration is given by a mixture of $\Psi(d^6) = c_1 * \Psi(HS) + c_2 * \Psi(LS)$, where $c_1$ and $c_2$ are numerical coefficients. The quantum spin fluctuations between HS and LS states (HS → LS, LS

→ HS) do not require any energy at the critical point since the HS and LS states have the same energy. We should emphasize here that these are fluctuations of the spin value contrary to the conventional fluctuations of the spin direction in the magnetic state. That is, the quantum spin fluctuations suppress dramatically the magnetic order at the critical point.

It has been shown in recent years that the quantum critical point phenomena appear in many quantum systems where the thermodynamic order parameter (here the sublattice magnetization in the antiferromagnetic HS state) disappears at T = 0 under some external influences (such as the high pressure variable investigated here). [30] The quantum phase transition has a Berry phase-like topological order parameter, [30,31] and the quantum fluctuations in the critical point are the electronic HS-LS fluctuations between two degenerate HS and LS terms.

The shaded area in Fig. 4 (red color) shows the region of the coexisting HS and LS states. The coexistence of the different spin states at finite temperatures has been observed experimentally in several $Fe^{3+}$-containing oxides, [29, 32, 33] and has been explained as the consequence of the thermal fluctuation effects between the electronic HS (S=5/2) and LS (S=1/2) $Fe^{3+}$ states. [29] Additionally, the coexistence of the HS and LS $Fe^{2+}$ irons has been found in $(Mg_{0.75},Fe_{0.25})O$ at temperatures between 300 K and 2000 K using X-ray emission spectroscopy in a laser-heated DAC. [34] As was shown previously [33], the width of the coexisting region depends strongly on temperature, and decreases dramatically as the temperature approaches zero due to the suppression of the thermal fluctuations.

### 3.4. Magnetic properties of ferropericlase at high pressures and temperatures relevant to the Earth's lower-mantle conditions

Since the spin transition is of great interest to the deep-Earth scientists, here we have used theoretical calculations to extend the experimental results to high pressures and temperatures relevant to the lower-mantle conditions [35]. The magnetic properties of ferropericlase have been analyzed above the critical pressure $P_c$ using the multi-electron generalized tight binding (GTB) approach [36, 37]. In the $d^6$ state of $Fe^{2+}$, the critical pressure is determined by the crossover of the HS ($^5T_2$) and LS ($^1A_1$) terms with the energies [38]

$$E_{HS}\left(d^6, {}^5T_2\right) = 6\varepsilon_d + 15A - 21B - 4Dq, \qquad (2)$$

$$E_{LS}\left(d^6, {}^1A_1\right) = 6\varepsilon_d + 15A - 16B + 8C - 24Dq, \qquad (3)$$

where $\varepsilon_d$ is a single $d$-electron atomic energy, $A$, $B$, and $C$ are the Racah parameters, and $10Dq$ is the crystal field splitting energy in the cubic system. With increasing pressure, the crystal field splitting energy increases as $10Dq(P) = 10Dq + \alpha P$. All other parameters are intra-atomic and thus are pressure-independent. The critical pressure $P_c$ can thus be represented as $(2.5 B + 4 C - 10 Dq)/\alpha$.

In the present calculations, we use the Racah parameters $A = 2$ eV, $B = 0.084$ eV and $C = 0.39$ eV, obtained for $Fe^{3+}$ ions in $FeBO_3$, [38] since the difference in the parameter values for $Fe^{2+}$ and $Fe^{3+}$ has been shown to be negligible. [39] The crystal field splitting energy is material dependent, and is equal to 10800 cm$^{-1}$ or 1.34 eV for ferropericlase at ambient pressure. [13] Using these parameters and the experimentally determined $P_c = 56$ GPa, we find $\alpha = 0.0078$ eV/GPa.

At pressures above the critical point one can then estimate the value of the spin gap $\varepsilon_S$ using the equation

$$\varepsilon_S = E_{HS}\left(d^6, {}^5T_2\right) - E_{LS}\left(d^6, {}^1A_1\right) + 2\alpha P = 2(10Dq - 2.5B - 4C + \alpha P) = 2\alpha(P - P_c). \qquad (4)$$

This gap increases with pressure (see Table I). At a fixed pressure above $P_c$, it allows determination of the concentration of the thermally excited HS $Fe^{2+}$ state; however, this concentration is also temperature-dependent. At low temperatures, all $Fe^{2+}$ ions are in the LS state at $P > P_c$, and magnetization is zero. At high temperatures, the magnetic moment may appear from the thermo-activated HS $Fe^{2+}$ ions.

Similar electronic structures and magnetic properties have been found in $LaCoO_3$ at ambient pressure. Indeed, the $Co^{3+}(d^6)$ ion has the LS ($^1A_1$) ground state with a small spin gap $\varepsilon_S \approx 150$ K. [40, 41] The LDA+GTB calculations of $LaCoO_3$ (Ref. [42]) have confirmed the known temperature-dependent magnetic susceptibility $\chi$ with the maximum at $T_S = \varepsilon_S/k_B$ ($k_B$ is the Boltzman constant). The same behavior in the temperature-dependent susceptibility is also expected for the LS ferropericlase (Fig. 5). The $\chi(T)$ dependence is determined by the competition between two major factors: (i) At low temperatures ($T \ll T_S$), the small admixture of the excited HS term into the nonmagnetic LS ($S=0$) state results in an increase in the magnetization and susceptibility $\chi(T)$. (ii) At high temperatures ($T \gg T_S$), the $\chi(T)$ value, generated by the excited HS state, decreases due to the standard Curie law. These two opposite regimes co-exist at $T \sim T_S$ resulting in a $\chi(T)$ maximum (Fig. 5). In Table I, the values of $\varepsilon_S$ and corresponding temperatures $T_S$ are shown for each pressure in ferropericlase, according to: $k_B T_S = 2\alpha(P - P_c)$. The dashed line in the red shade area of the phase diagram in Fig. 4 indicates the P,T- line where the $\chi$ maximum is expected at the lower-mantle conditions.

Further consideration of the high-temperature spin crossovers of $Fe^{2+}$ in ferropericlase (Fig. 6) has revealed that the difference in the spin and orbital degeneracy of the HS and LS terms results in an asymmetry of the phase diagram as the fraction of HS and LS states varies with high P-T. The fraction of the HS state is given by

$$n_{HS}(P,T) = \frac{g_{HS}\exp(-E_{HS}/kT)}{g_{HS}\exp(-E_{HS}/kT)+g_{LS}\exp(-E_{LS}/kT)} = \frac{1}{1+\frac{g_{LS}}{g_{HS}}\exp\left(\frac{E_{HS}-E_{LS}}{kT}\right)}. \quad (5)$$

At $g_{HS} = g_{LS}$ the HS and LS (P,T)-distribution would be symmetrical relative to the $P_c(T)$-line. However, for the HS-$d^6$ state the spin and orbital values are $S = 2$, $L = 1$ leading to $g_{HS} = (2S+1)(2L+1) = 15$, whereas for the LS-$d^6$ state $S = 0$, $L = 0$ and $g_{LS} = 1$ (Fig. 6a). The sharp transformation of the HS state into the LS state at zero temperature reflects the effect of the quantum phase transition, which may be given analytically by the following equation

$$P(n_{HS}) = P_c + kT/(2\partial\varepsilon_S/\partial P)\times\ln(n_{LS}g_{HS}/n_{HS}g_{LS}), \quad (6)$$

here $P(n_{HS})$ is the pressure where the relative fraction of HS $Fe^{2+}$ ions is $n_{HS}$, whereas the fraction of the LS $Fe^{2+}$ ions is $n_{LS} = (1 - n_{HS})$. This equation describes each P-T point in the diagram, and relates particular fractions of the HS and LS states. The P - T relationship is shown to be linear for each given $n_{HS}$ value, thus a set of lines with different $n_{HS}$ begins in the quantum critical point (Fig. 6a).

For comparison, we plot in Fig. 6b the experimental data obtained for the same sample by the X-ray emission spectroscopy with laser heating [34]. General agreement of the calculated and measured diagrams is obvious. Some minor differences are in the behavior of lines presenting the constant values of different HS/LS fractions at $P > P_c$. The deviation may be explained by the contribution of high energy excited terms of the $Fe^{2+}$ ion, which may include the intermediate spin $S = 1$ state and the unit-cell-volume collapse in the LS state, which will reduce the HS component to a larger extent than the simple linear compressibility model predicts. Previously, similar distributions of the HS fraction in (P,T) plane have been obtained by mean-field [43] and LDA+U [44] calculations. The main novelty of the present consideration are: (i) the multielectron calculation of the $Fe^{2+}$ ion energies taking into account strong electron correlations; (ii) the unification of the continuous high temperature spin crossover with the sharp quantum phase transition at zero temperature, and (iii) calculations of magnetic susceptibility as function of pressure and temperature.

## 4. CONCLUSIONS

Our results show that the spin transition at the Earth's lower mantle conditions is a continuous crossover. At T = 2000 K, the smooth transformation of the spin state takes place at pressures between 50 and 90 GPa, which corresponds to the mid-lower mantle conditions over a depth of approximately 1000 km. These results on the quantum critical point, which exist at zero temperature, can be applied to understand the (P,T) phase diagram and properties of ferropericlase at relevant lower mantle conditions. Lower-mantle ferropericlase is expected to be subjected to pressures up to 136 GPa and temperatures as high as approximately 2800 K. Based on our experimental and theoretical results here, the lower-mantle pressure is expected to increase the spin gap and thus to increase the $T_S$ value. However, one should take into account the fact that this qualitative conclusion is only valid below the melting point of ferropericlase in the LS state at $P > P_c$. The melting point of pure MgO is approximately 3000-3400 K at ambient pressure [45,46] and increases to about 4000 K [45] or even to 6000 K [46,47] with increasing pressure to 50 GPa. In this case our model predicts that ferropericlase can exist in a paramagnetic state even at the middle layer of the lower mantle at depths between 1300 and 1900 km. It is conceivable that the occurrence of the paramagnetic ferropericlase with distinct physical and chemical properties can significant influence our understanding of the geophysics, geochemistry, and geodynamics of the planet's interior.

## ACKNOWLEDGMENTS

We thank Dr. Yu.S. Orlov and A. Wheat for useful discussion. This work is supported by the Russian Foundation for Basic Research grants #11-02-00636, #09-02-00171, #10-02-00251, #09-02-01527, #11-02-00291, the integration Grant #40 of SBRAS, the Presidium RAS program #18.7, and by grant of the Russian Ministry of Science 16.518.11.7021. This work at the UT Austin was supported by the US National Science Foundation (EAR-0838221), Energy Frontier Research in Extreme Environments Center, and the Carnegie/DOE Alliance Center. The support from the DOE grant DE-FG02-02ER45955 for the work at GL and at the APS synchrotron facility is greatly acknowledged. The synchrotron Mössbauer work was performed at HPCAT (Sector 16), APS, ANL. HPCAT is supported by DOE-BES, DOE-NNSA, NSF, and the W.M. Keck Foundation. APS is supported by DOE-BES, under Contract No. DE-AC02-06CH11357.

# REFERENCES

1. D. Y. Pushcharovskii, Phys. Usp. **45**, 439 (2002).

2. R. Jeanloz and E. Knittle, Philos. Trans. R. Soc. London Ser. A **328**, 377 (1989).

3. D. M. Sherman, J. Geophys. Res. **96**, 14299 (1991).

4. D. Y. Pushcharovskii, Rus. J. Nature (Priroda) **11**, 119 (1980).

5. K. K. M. Lee, B. O'Neill, W. R. Panero, S.-H. Shim, L. R. Benedetti, and R. Jeanloz, Earth Planet. Sci. Lett. **223**, 381 (2004).

6. R. I. Chalabov, I. S. Lyubutin, Z. I. Zhmurova, A. P. Dodokin, and T. V. Dmitrieva, Sov. Phys. Crystallography **27**, 312 (1982).

7. J. F. Lin, A. G. Gavriliuk, V. V. Struzhkin, S. D. Jacobsen, W. Sturhahn, M. Y. Hu, P. Chow, and C.-S. Yoo, Phys. Rev. B **73**, 113107 (2006).

8. I. Y. Kantor, L. S. Dubrovinsky, and C. A. McCammon, Phys. Rev. B **73**, 100101 (2006).

9. A. G. Gavriliuk, J. F. Lin, I. S. Lyubutin, and V. V. Struzhkin, JETP Lett. **84**, 161 (2006).

10. J. F. Lin, V. V. Struzhkin, A. G. Gavriliuk, and I. S. Lyubutin, Phys. Rev. B **75**, 177102 (2007).

11. J. Li, in *In Post-Perovskite: The Last Mantle Phase Transition* edited by J. B. K. Hirose, T. Lay, and D. Yuen (American Geophysical Union, Washington, DC, 2007), p. 47.

12. J. F. Lin, S. D. Jacobsen, and R. M. Wentzcovitch, Eos, Transaction, American Geophysical Union **88**, 13 (2007).

13. J. F. Lin, A. G. Gavriliuk, W. Sturhahn, S. D. Jacobsen, J. Zhao, M. Lerche, and M. Hu, American Mineralogist **94**, 594 (2009).

14. I. S. Lyubutin, A. G. Gavriliuk, K. V. Frolov, J. F. Lin, and I. A. Trojan, JETP Lett. **90**, 617 (2009).

15. A. F. Goncharov, V. V. Struzhkin, and S. D. Jacobsen, Science **312**, 1205 (2006).

16. H. Keppler, I. Kantor, and L. S. Dubrovinsky, American Mineralogist **92**, 433 (2007).

17      J. F. Lin, S. T. Weir, D. D. Jackson, W. J. Evans, and C. S. Yoo, Geophysical Research Letters **34**, L16305 (2007).

18      J. F. Lin, V. V. Struzhkin, S. D. Jacobsen, M. Hu, P. Chow, J. Kung, H. Liu, H. K. Mao, and R. J. Hemley, Nature **436**, 377 (2005).

19      Y. Fei, L. Zhang, A. Corgne, H. C. Watson, A. Ricolleau, Y. Meng, and V. B. Prakapenka, Geophysical Research Letters **34**, L17307 (2007).

20      J. F. Lin, W. Sturhahn, J. Zhao, G. Shen, H. K. Mao, and R. J. Hemley, Science **308**, 1892 (2005).

21      S. Speziale, A. Milner, V. E. Lee, S. M. Clark, M. Pasternak, and R. Jeanloz, Proc. Nat. Acad. Sci. USA **102**, 17918 (2005).

22      D. P. Dobson, N. S. Cohen, Q. A. Pankhurst, and J. P. Brotholt, American Mineralogist **83**, 794 (1998).

23      A. G. Gavriliuk, A. A. Mironovich, and V. V. Struzhkin, Review of Scientific Instruments **80**, 043906 (2009).

24      Y. V. Shvyd'ko, Phys. Rev. B **59**, 9132 (1999).

25      G. K. Wertheim ed., *Mossbauer Effect. Principles and Applications.* (Academic Press, New York and London, 1964).

26      W. L. Roth, Acta Crystallogr. **13**, 140 (1960).

27      J. M. Ziman, *Models of Disorder* (Cambridge University Press, Cambridge, London, New York, Melbourne, 1979).

28      E. Cotton, J. Phys. Radium **21**, 265 (1960).

29      I. S. Lyubutin, A. G. Gavriliuk, V. V. Struzhkin, S. G.Ovchinnikov, S. A. Kharlamova, L. N. Bezmaternykh, and M. Y. Hu, JETP Lett. **84**, 518 (2006).

30      S. Sachdev, *Quantum phase transitions* (Cambridge University Press, Cambridge, 2001).

31      A. I. Nesterov and S. G. Ovchinnikov, JETP Lett. **90**, 580 (2009).

32      I. S. Lyubutin and A. G. Gavriliuk, Physics – Uspekhi **52**, 989 (2009).

33  A. G. Gavriliuk, V. V. Struzhkin, I. S. Lyubutin, S. G.Ovchinnikov, M. Y. Hu, and P. Chow, Phys. Rev. B **77**, 155112 (2008).

34  J. F. Lin, G. Vankó, S. D. Jacobsen, V. Iota, V. V. Struzhkin, V. B. Prakapenka, A. Kuznetsov, and C.-S. Yoo, Science **317**, 1740 (2007).

35  S. G. Ovchinnikov, JETP Lett. **94**, 192 (2011).

36  S. G. Ovchinnikov, JETP Lett. **77**, 676 (2003).

37  S. G. Ovchinnikov, J. Phys.: Condens. Matter **17**, S743 (2005).

38  A. M. Gavrilyuk, I. A.Trojan, S. G. Ovchinnikov, I. S. Lyubutin, and V. A. Sarkisyan, JETP **99**, 566 (2004).

39  Y. Tanabe and S. Sugano, J. Phys. Soc. Jpn. **9**, 766 (1954).

40  Z. Ropka and R. J. Radwanski, Phys. Rev. B **67**, 172401 (2003).

41  M. W. Haverkort, et al., Phys. Rev. Lett. **97**, 176405 (2006).

42  S. G. Ovchinnikov, Y. S. Orlov, I. A. Nekrasov, and Z. V. Pchelkina, arXiv 1005.1732 v.1 (2010).

43  W. Sturhahn, J. M. Jackson, and J. F. Lin, Geophysical Research Letters **32**, L12307 (2005).

44  T. Tsuchiya, R. M. Wentzcovitch, C. R. S. daSilva, and S. deGironcoli, Phys. Rev. Lett. **96**, 198501 (2006).

45  A. Zerr and R. Boehler, Nature **371**, 506 (1994).

46  L. Vocadlo and G. D. Price, Phys. Chem. Miner. **23**, 46 (1996).

47  R. E. Cohen and Z. Gong, Phys. Rev. B **50**, 12301 (1994).

**Figure captions**

**Fig. 1.** Representative low-temperature transmission $^{57}$Fe-Mössbauer spectra at ambient pressure. **(a)**, $(Mg_{0.75},Fe_{0.25})O$; **(b)**, $(Mg_{0.8},Fe_{0.2})O$. Temperature-dependent average values of magnetic hyperfine field $<H_{hf}>$ at iron nuclei obtained from the spectra of $(Mg_{0.75},Fe_{0.25})O$ **(c)** and $(Mg_{0.8},Fe_{0.2})O$ **(d)**. Solid and dashed lines are guides to the eye.

**Fig. 2.** Temperature-dependent quadrupole splitting parameter ($\varepsilon$) obtained from the $^{57}$Fe-Mössbauer spectra in the paramagnetic region of $(Mg_{0.75},Fe_{0.25})O$ (brown color) and $(Mg_{0.8},Fe_{0.2})O$ (blue color) [$\varepsilon = e^2qQ/2$, where $Q$ is the nuclear quadrupole moment and $eq = V_{zz} = \partial^2 V/\partial z^2$ is the electric field gradient]. Solid lines are guides to the eye.

**Fig. 3.** Representative low-temperature synchrotron Mössbauer spectra of $(Mg_{0.75},Fe_{0.25})O$ at 38 GPa **(a)** and 55 GPa **(b)**. High-frequency quantum beats indicate a magnetic ordering of $Fe^{2+}$ ions in the HS state, whereas the low-frequency quantum beats indicate the paramagnetic state of $Fe^{2+}$ ions in the HS state. Absence of the quantum beats indicates the occurrence of the diamagnetic state of the LS $Fe^{2+}$ ions. Solid lines are calculated data.

**Fig. 4.** Magnetic phase diagram of ferropericlase $(Mg_{0.75},Fe_{0.25})O$ at high pressures and low temperatures. The dashed blue line separates the regions of the HS paramagnetic and antiferromagnetic phases between 0 and 50 GPa. Above 56 GPa, the diagram shows the LS diamagnetic phase which appears due to the HS → LS crossover. A quantum critical point appears in the diagram at T = 0 and $P = P_c$ which can be explained in terms of the geometric phase of the topological origin that is considered as an order parameter in the spin crossover phenomena [36].

In the shade (red) area, the HS and LS states coexist due to the thermal fluctuations between the electronic HS and LS states. In the upper part of this area (shaded red), the region of the coexisting HS and LS states is extended to high temperatures relevant to the lower mantle conditions (calculated from the Ref. [34] experimental data). The dash line in this area indicates the position of maximum of the magnetic susceptibility $\chi$ of ferropericlase.

**Fig. 5**. The expected temperature dependence of the magnetic susceptibility above the critical pressure $P_c$ of the HS-LS crossover at the Earth's lower mantle conditions. $T_S$ is the spin gap value (in the temperature scale) which is proportional to $(P - P_c)$. At $T \ll T_S$, the admixture of excited HS term into the nonmagnetic LS state results in an increase of magnetization and susceptibility. At $T \gg T_S$, the $\chi$ value, generated by the excited HS state, decreases due to the standard Curie law. These two opposite regimes result in a maximum of $\chi$ at $T \sim T_S$.

**Fig. 6**. Phase diagram of spin crossover of $Fe^{2+}$ in $(Mg_{0.75},Fe_{0.25})O$ calculated from Eq. (6) (a) and measured in [34] by X-ray emission spectroscopy with laser heating (b). Colors in the vertical columns on the right represent fractions of the high-spin iron $n_{HS}$.

**Tables**

**Table I.** The pressure dependence of the spin gap $\varepsilon_S$ in ferropericlase above the critical pressure $P_c$ of the HS-LS crossover. $T_s$ is the temperature were a maximum of magnetic susceptibility is expected.

| $(P - P_c)$, GPa | 5 | 10 | 15 | 20 |
|---|---|---|---|---|
| $\varepsilon_S$, eV | 0.078 | 0.156 | 0.234 | 0.312 |
| $T_S$, K | 906 | 1812 | 2721 | 3624 |

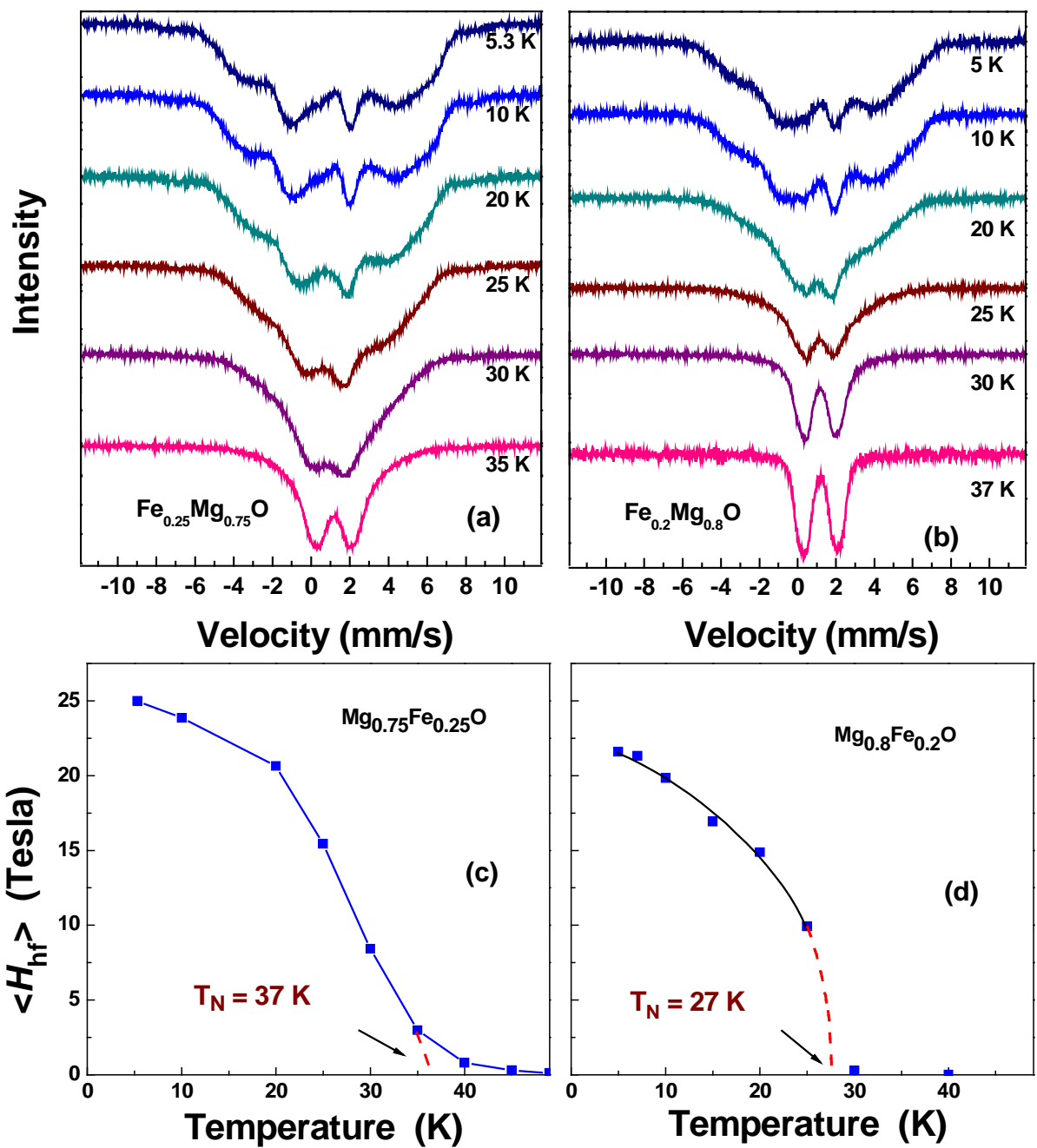

Figure 1.

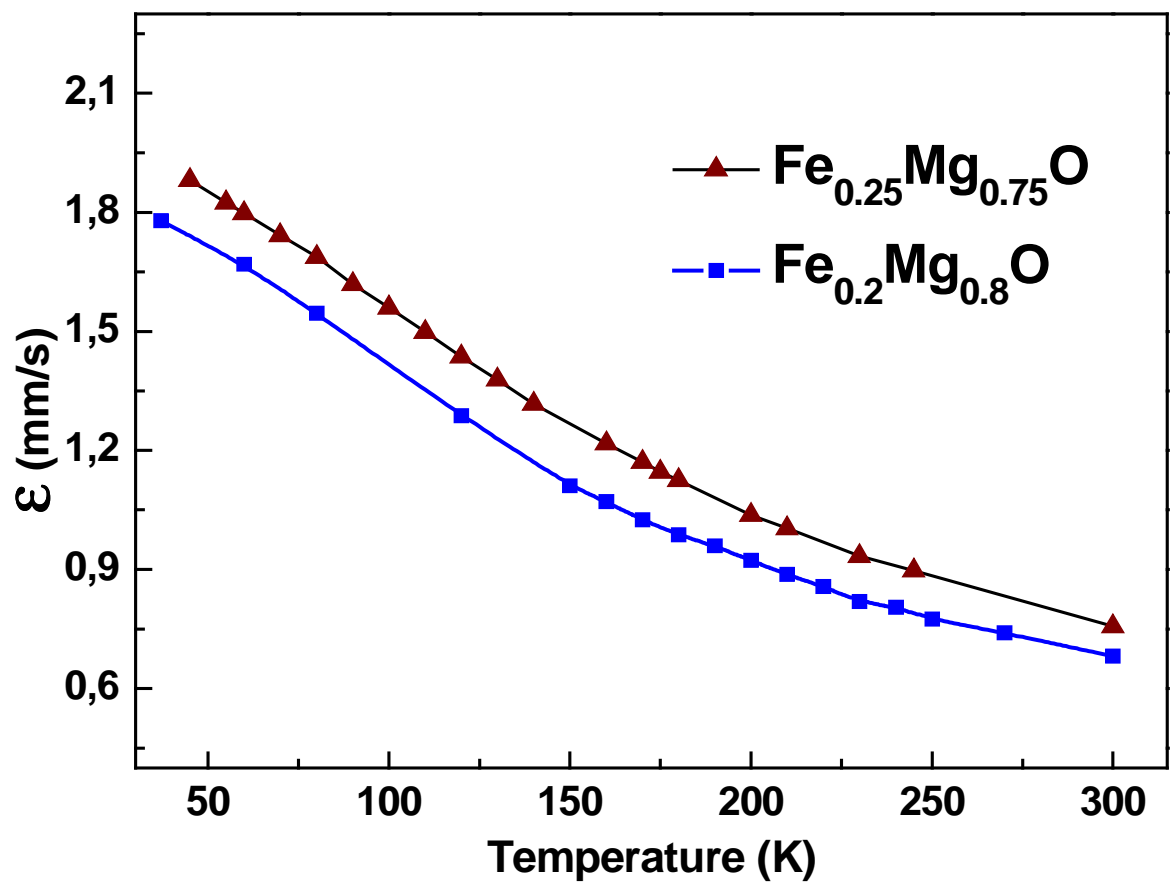

Figure 2.

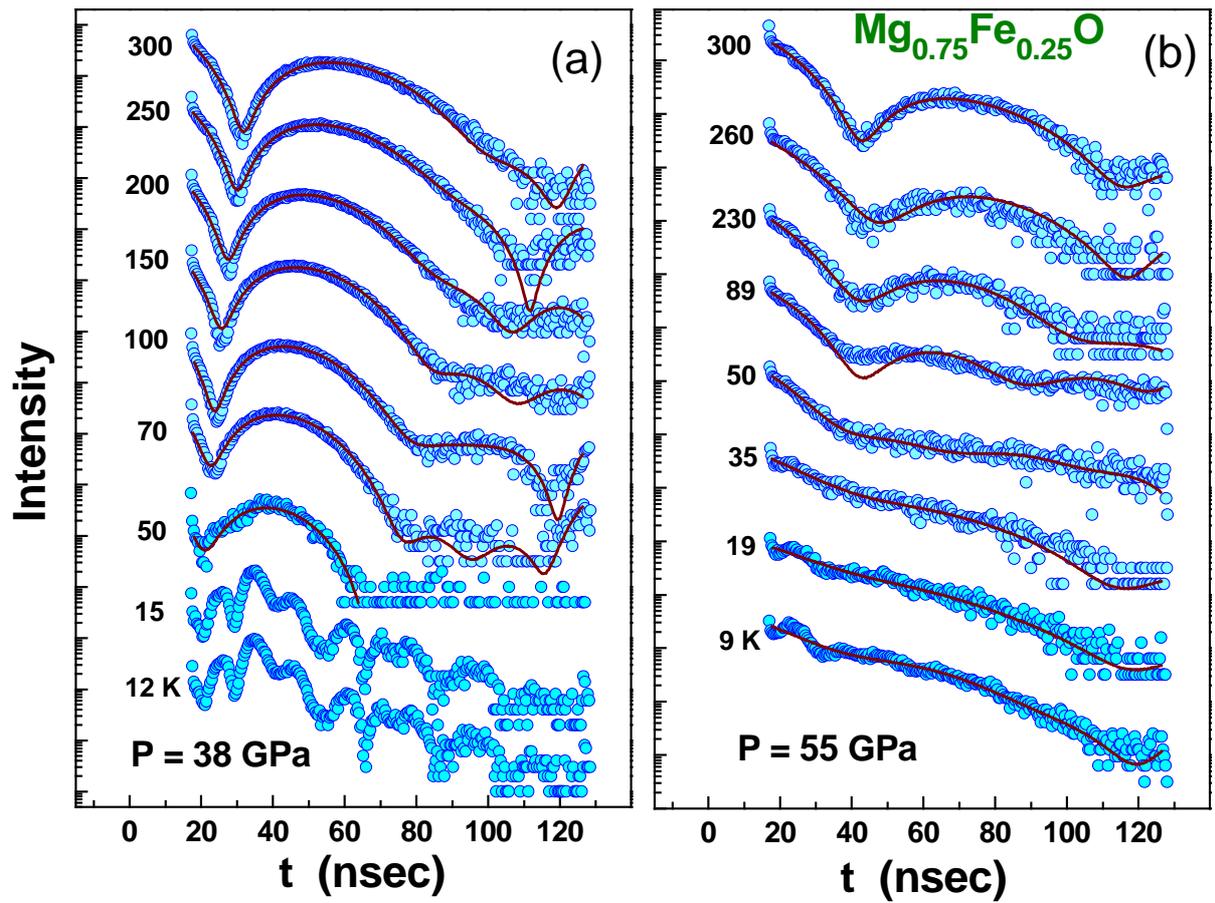

Figure 3.

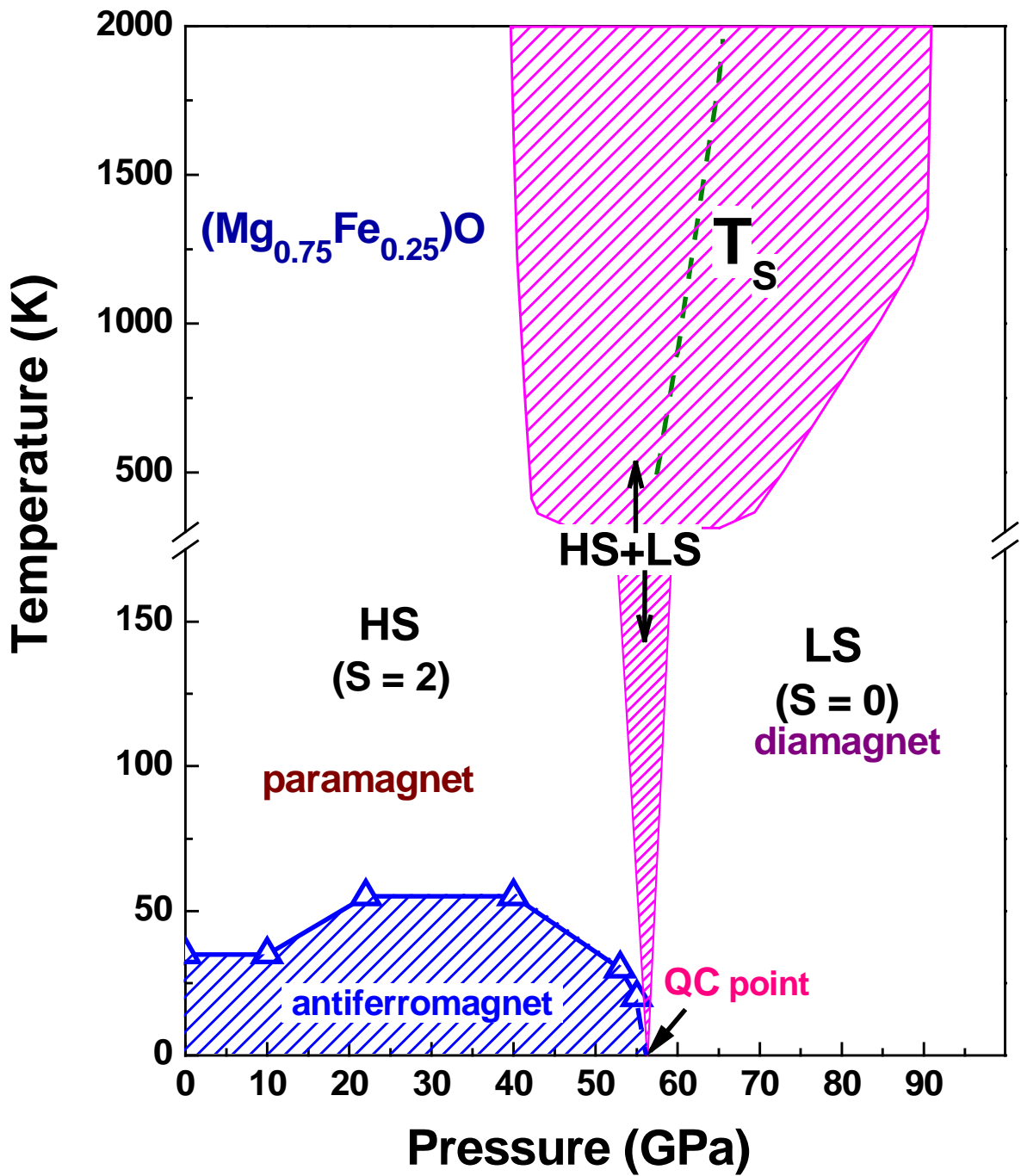

Figure 4.

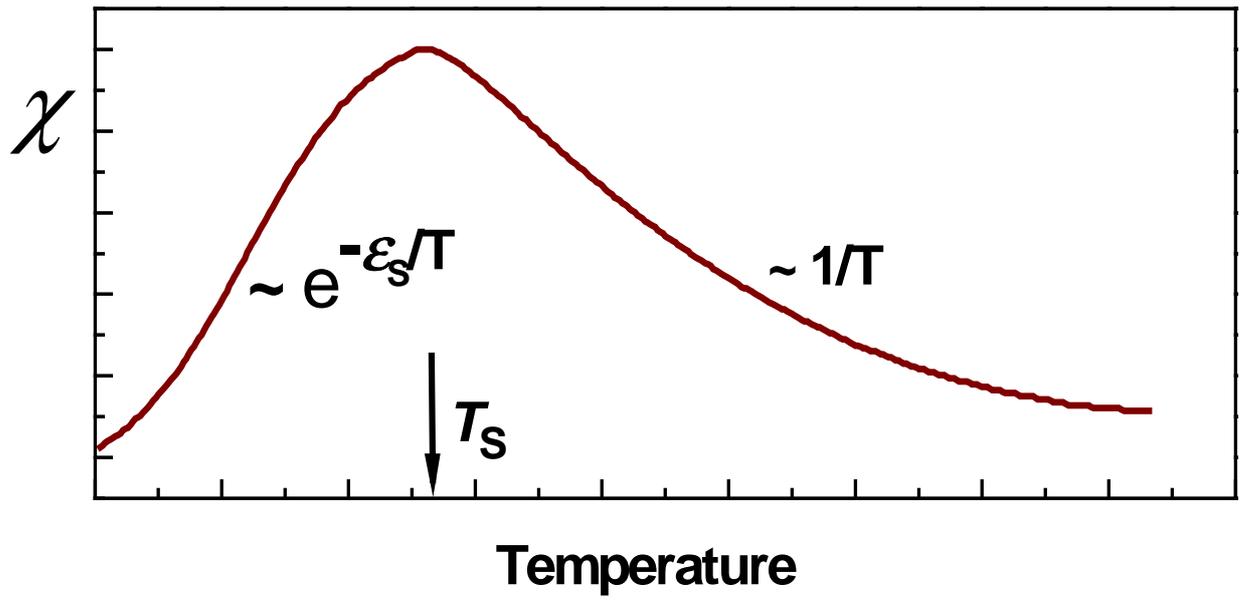

Figure 5.

Figure 6.

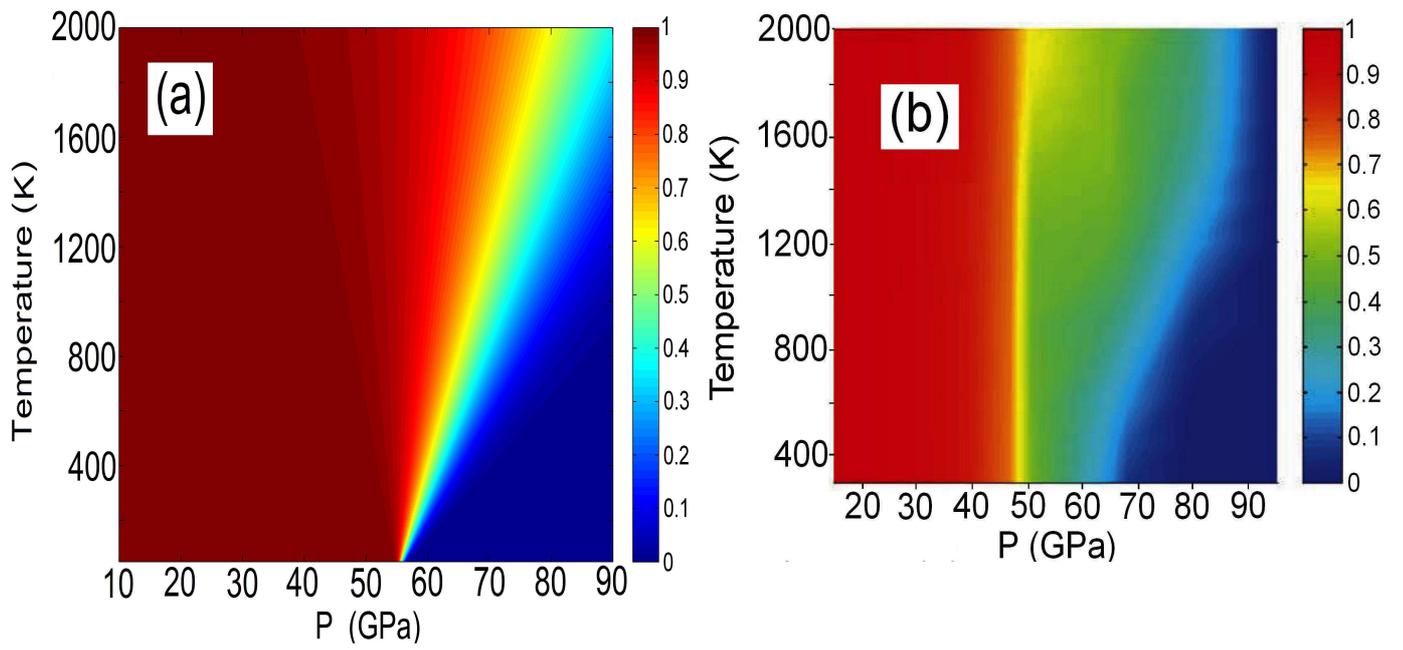